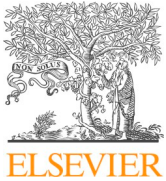

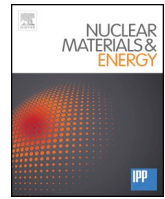

# Neutronic study of UO₂-BeO fuel with various claddings

Shengli Chen[1], Cenxi Yuan[*]

*Sino-French Institute of Nuclear Engineering and Technology, Sun Yat-sen University, Zhuhai, Guangdong 519082, China*

ABSTRACT

The neutronic properties of UO₂-BeO fuel with various claddings are investigated through the Monte Carlo method and the Linear Reactivity Model. A second order polynomial function is suggested to describe the relationship between the difference of the reactivity at the End of Cycle and two factors, the uranium enrichment and the volume fraction of BeO in the UO₂-BeO fuel. The uranium enrichment is determined to ensure the same cycle length for each BeO fraction with zircaloy, FeCrAl, and SiC claddings. Similar neutronic properties are observed between the zircaloy and SiC claddings for a given BeO fraction, including the infinite multiplication factor, the gas release, the power distribution, and the isotopic concentrations. An important feature of UO₂-BeO fuel is the production of ⁴He, which is significant compared with current UO₂ fuel. Lower production rates of xenon and krypton are found in the case of the UO₂-BeO-FeCrAl fuel-cladding system, while those in the UO₂-BeO-zircaloy and UO₂-BeO-SiC systems are quite similar to the current UO₂-zircaloy system. The power distribution in an assembly and in a fuel pellet is flatter for a higher BeO fraction with the same cladding. Higher peak power is found in fuel assemblies with FeCrAl compared to the other two claddings, while the radial power distribution is quite similar for the three claddings.

## 1. Introduction

After the Fukushima Daiichi nuclear disaster in 2011, extensive studies on the Beyond Design Basis Accident (BDBA) have been performed. The US Department of Energy (DOE) has proposed the concept of Accident Tolerant Fuel (ATF) for advanced nuclear fuel and cladding options. Three approaches of the ATF concept have been suggested in 2014 as [1]:

(1) Replacement or improvement of the current UO₂ ceramic fuel;
(2) Improvement of the oxidation resistance for the cladding by modifying the current zircaloy alloy;
(3) Utilization of new high oxidation resistance cladding materials.

Many potential ATFs and claddings have been proposed by the US DOE Office of Nuclear Energy Advanced Fuels Campaign [2]. A test of ATF assemblies in an operating commercial reactor recently started in the Edwin 1 Hatch nuclear power plant [3]. A large safety margin to the melting of nuclear fuel can increase the accident tolerance. High thermal conductivity and high melting temperature of the fuel are thus two important factors. Contrary to the most recently proposed candidate ATFs in the past decade such as the high uranium density fuel

U₃Si₂ [4–7], studies on UO₂-BeO fuel can be traced back to 1963 [8]. Loading of BeO into ceramic UO₂ fuel can improve the thermal conductivity due to the much higher thermal conductivity of BeO compared to UO₂. As shown in Fig. 1, the thermal conductivity of the BeO (interpolated by Zhou [9] according to the experimental data from Ref. [10]) is 22 times larger than that of UO₂ at room temperature. At the temperature of 1500 K, the ratio of the thermal conductivity of the BeO to that of the UO₂ is 3 and 7 when the Fink model [11] and the Harding and Martin (HM) formula [12] are used, respectively.

Many investigations have been performed on UO₂-BeO fuel with different BeO volume fractions. The irradiation behavior of 12% ²³⁵U enriched UO₂-(30vol%)BeO fuel was tested by Johnson and Mills in 1963 [8]. Ishimoto et al. experimentally showed the large improvement of the thermal conductivity by adding 1.1, 2.1, 3.2, 4.2 and 36.4 vol% BeO in UO₂ fuel [13]. Ref. [13] shows that the thermal conductivities of UO₂-(4.2vol%)BeO fuel at 1100 K are higher than that of UO₂ by 10% or 25% depending on the incorporation method. Later, Li et al. investigated the thermal conductivity of UO₂-BeO fuel using statistical continuum mechanics for BeO fraction from 0 to 100% [14]. Chandramouli and Revankar developed a thermal model of UO₂-BeO fuel and analyzed the performance during a Loss of Coolant Accident (LOCA) [15]. Some neutronic properties of the UO₂-(5vol%)BeO and

* Corresponding author.
*E-mail address:* yuancx@mail.sysu.edu.cn (C. Yuan).
[1] Current address at: CEA, Cadarache, DEN/DER/SPRC/LEPh, Saint Paul Les Durance 13108, France.

**Nomenclature**

| | |
|---|---|
| $k_{inf}$ | infinite multiplication factor |
| $b$ | batch number of fuel assembly in the reactor |
| $e_b$ | fuel exposure of the batch $b$ (EFPD) |
| $P_b$ | relative assembly power of the batch $b$ |
| $V_b$ | number of the assemblies for the batch $b$ |
| $\Delta k_{core}$ | difference on $k_{inf}$ to the reference case in the whole core level |
| $k_{inf,\,b}$ | $k_{inf}$ of the batch $b$ |
| $k_{inf,b}^{ref}$ | $k_{inf}$ of the batch $b$ for the reference case |
| $\Delta k_{inf,\,b}$ | difference on $k_{inf}$ to the reference case for the batch $b$ |
| $x$ | volume fraction of BeO in the $UO_2$-BeO fuel (%) |
| $y$ | $^{235}U$ enrichment (%) |
| $y_c$ | critical $^{235}U$ enrichment (%) by keeping the cycle length according to the LRM |
| $y_U$ | $^{235}U$ enrichment by keeping the $^{235}U$ loading in the fuel |
| $R^2$ | the coefficient of determination of least square fitting |

$UO_2$-(10vol%)BeO fuel have been studied by Smith [16]. McCoy and Mays showed that the incorporation of BeO (4vol% and 9.6vol%) can reduce the internal rod pressures, and fission gas release [17]. Zhang et al. investigated the neutronic properties of $UO_2$-BeO fuel with 5vol%, 10vol%, and 30vol%BeO with the critical fuel enrichments determined by the reactivity of fuel assembly [18].

Recently, Zhou and Zhou analyzed the thermophysical and mechanical properties of the $UO_2$-(36.4vol%)BeO fuel with different claddings [9]. The present work focuses on the neutronic behaviors of the $UO_2$-BeO fuel with different volume fractions of BeO (up to 36.4vol %) in combination with the different claddings in Pressurized Water Reactor (PWR) conditions using the Monte Carlo code RMC [19]. The critical fuel enrichments are determined by the Linear Reactivity Model (LRM) [20], which is summarized in Section 2. Zircaloy-4 cladding is studied in the present work because it is widely used in current PWRs. The FeCrAl cladding has better oxidation resistance than the current zircaloy cladding [1]. The ceramic SiC is a candidate ATF cladding material because of its excellent oxidation resistance in the environment of high temperature steam [21]. The maximum service temperature of SiC can be 900 °C, while that of the zircaloy alloy is only 400 °C [22]. Katoh et al. proved its stability at high neutron fluence [23]. Therefore, the present study includes the zircaloy, FeCrAl, and SiC claddings.

An important feature of $^9Be$, 100% abundance in natural beryllium (Be), is the relatively large (n,α) cross section at the neutron energies below 3 MeV as shown in Fig. 2. No information on the (n,α) cross sections of $^{235}U$ or $^{238}U$ is given in the ENDF/B-VII.0 nuclear data library [24], which is used in our simulations. The quite small (n,α) cross sections of fissionable nuclei have been evaluated in recently released nuclear libraries, including ENDF/B-VIII.0 [25] and JEFF-3.3 [26]. In addition to direct $^4He$ production via (n,α) reaction of $^9Be$, large thermal (n,t) cross section of $^6Li$, which is from a beta decay of $^6He$ (product of $^9Be$ (n,α) reaction) with 0.82 s half-life, leads to both tritium and $^4He$ gas production. Regardless of fission products, the main production of the helium should come from $^9Be$. Either the produced gas can lead to swelling of the fuel pellet or, if released from the fuel, cause increased pressure in the free volume of the fuel rod. A potential consequence could be fuel cladding mechanical interaction or even cladding damage. The study of the gas production induced by the loading of $^9Be$ is thus necessary.

The hotspots in a reactor must be identified, as they are limiting with respect to safety criteria such as critical heat flux ratio or fuel melting. The peak power in an assembly is analyzed for each BeO fraction and each cladding. The periphery phenomenon in a fuel rod has been observed in PWR [27,28] due to the spatial self-shielding. Studies on the radial distributions of the physical properties are of interest for the new types of fuels. The radial distributions of the power and isotopic concentrations of $^{235}U$ and $^{239}Pu$ are investigated.

Section 2 presents the methods used in the present study, such as the simulation methods and the treatment of the results. Section 3 summarizes the simulation results and the corresponding discussions, including critical fuel enrichment for each volume fraction of BeO, the gas production, the peak power in each assembly, and the radial properties of the critical compositions determined in the present work. The last section points out the main conclusions of the present work.

## 2. Methods

The present study is carried out for a typical 17 by 17 fuel assembly for a PWR, except that the investigations of radial distributions of power and isotopic concentration in a fuel pellet are based on the geometry shown in Ref. [29]. The geometrical parameters and physical conditions for simulation are detailed in our previous works [29–31]. The main parameters used in the present study are summarized in Table 1. A typical cladding thickness of 572 μm is used for the current Zr-4 cladding and the SiC cladding. It is noteworthy that even if thicker SiC cladding would be used, the results are not so sensitive to the thickness of SiC cladding because of its quite small thermal neutron cross sections. Due to the better mechanical properties of stainless steel compared to the zircaloy alloy, a lower thickness of 350 μm was proposed for stainless steel claddings for Light Water Reactor (LWR) fuel [32]. The present work uses 350 μm thickness for the FeCrAl cladding.

In order to keep the thermohydraulic properties constant, the same outer cladding radius is chosen for all cases. Since different thicknesses are used for different claddings, the corresponding radii of fuel pellets are different. Table 2 summarizes the radii and relative volumes of different claddings. It is noteworthy that a thickness of 350 μm leads to about 11% higher fuel volume compared to the reference case. The reference case referred in this paper is a current $UO_2$-Zr-4 cladding with 4.9% $^{235}U$ enrichment, 572 μm cladding thickness, and 38.33 MW/tU specific power density.

In the present work, all the neutronic simulations are performed with the 3D Monte Carlo based code RMC [19] using the ENDF/B-VII.0 nuclear data library [24]. The simulations carried out in the present study are performed with uniform temperature for each composition. The corresponding temperatures are given in Table 1. It is noteworthy that the same fuel temperature is used for all simulations in the present work, even if a higher BeO incorporation leads to a lower average fuel temperature. The steps used in depletion calculations are the same as the steps used in Ref. [29].

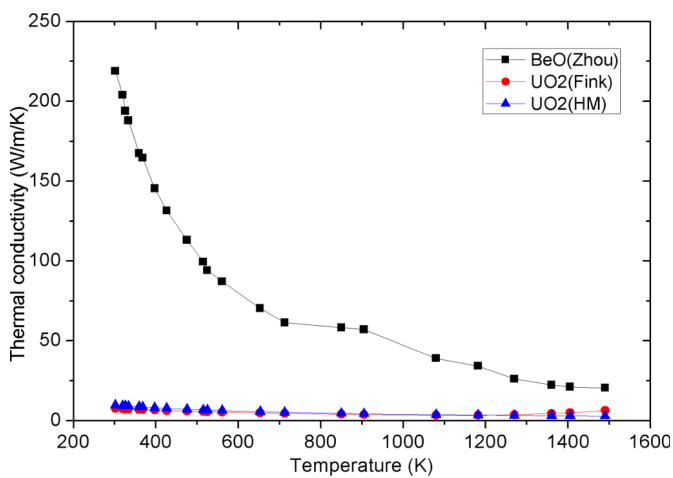

**Fig. 1.** The thermal conductivity of BeO and $UO_2$.

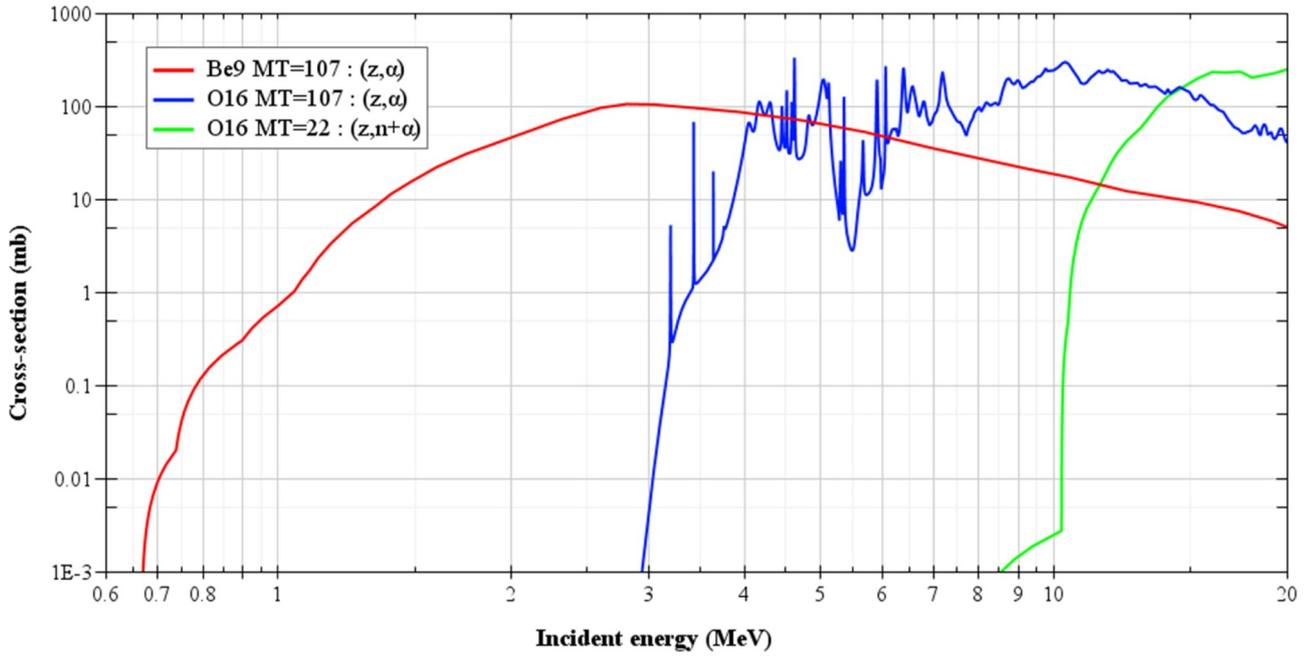

**Fig. 2.** Gas production cross section of $^9$Be and $^{16}$O in ENDF/B-VII.0 [24]. No (n,α) cross section of $^{235}$U or $^{238}$U exists in ENDF/B-VII.0.

**Table 1**
Key parameters of special fuel assembly designs.

| Property | Unit | Value |
|---|---|---|
| Zr-4 composition | wt% | Fe/Cr/Zr/Sn = 0.15/0.1/98.26/1.49 |
| FeCrAl composition | wt% | Fe/Cr/Al = 75/20/5 |
| SiC composition | at% | Si/C = 1/1 |
| Cladding density | g/cm³ | 6.56 (Zr-4); 7.10 (FeCrAl); 2.58 (SiC) |
| Cladding outer radius | mm | 4.75 |
| Cladding thickness | μm | 350 (FeCrAl); 572 (Zr-4, SiC) |
| Gap thickness | μm | 83 |
| Fuel density | g/cm³ | 10.47 (UO₂); 2.99 (BeO) |
| $^{235}$U enrichment | % | 4.9[a] |
| Average fuel temperature | K | 900[b] |
| Coolant temperature | K | 580 |
| Cladding and gap temperature | K | 600 |
| Boundary condition for simulations | - | Reflective |

[a] 4.9% $^{235}$U enrichment is used for the reference case. Fuel enrichments of other cases are detailly given in Section 3.1.

[b] Even if the average fuel temperature is lower for a higher BeO incorporation, 900 K is used in the present work for all simulations.

**Table 2**
Geometrical characters of fuel rods.

| Cladding | Cladding thickness (μm) | Radius of fuel pellet (mm) | Ratio of volume to the reference case |
|---|---|---|---|
| Zr-4 and SiC | 572 | 4.261 | 1.000 |
| FeCrAl | 350 | 4.483 | 1.107 |

Since all simulations are performed with RMC neutron transport code, the present study computes the gas production without considering diffusion and release of gas. For studying the radial distributions of power and isotopic concentrations, fuel pellet is divided into nine concentric rings with more dense points near the surface of the fuel pellet to treat the periphery phenomenon. The explanations and verification of such division are given in Ref. [29].

The LRM is used to calculate the equivalent reactivity at the End of Cycle (EOC) of the core by using the infinite multiplication factor $k_{inf}$ calculated in an assembly. In comparison to the reference case, the LRM points out that the difference on $k_{inf}$ in whole core level ($\Delta k_{core}$) is the weighted average value of difference on $k_{inf}$ of assemblies with different batches [20]:
Eq. (1).

$$\Delta k_{core} = \frac{\sum_b \Delta k_{inf,b}(e_b)P_b V_b}{\sum_b P_b V_b},\tag{1}$$

where $\Delta k_{inf,b} = k_{inf,b} - k_{inf,b}^{ref}$ is the difference of $k_{inf}$ between the treated fuel-cladding system $k_{inf,b}$ and the reference case $k_{inf,b}^{ref}$ for the batch $b$, $e_b$ is the fuel exposure of the batch $b$ and the values at the EOC in a typical Westinghouse PWR reactor are listed in Table 3 with the unit of Effective Full Power Day (EFPD), $P_b$ is the approximative contribution to the average thermal power in the core for the batch $b$, $V_b$ refers to the core volume fraction (in %) of the assemblies of batch $b$. The critical fuel enrichment can be determined via $\Delta k_{core} = 0$.

In this paper, the unit EFPD is used because it directly expresses fuel exposure by the time of full power operating of a reactor. The relationship between EFPD and burnup (noted by bu) is bu = EFPD × SPD, where SPD is the Specific Power Density (in MW/tU). SPD = 38.33 MW/tU for the reference case. In order to have the same power for different cases, the thermal power in a fuel pin SPD × $\rho_U V_U$ (where $\rho_U V_U$ is the mass of uranium) is the same for each case. Therefore, bu∝EFPD/$V_U$. For a given BeO volume fraction UO₂-BeO fuel, the burnup level of a fuel rod with 350 μm thick FeCrAl cladding is 0.903 times the burnup with 572 μm thick Zr-4 and SiC claddings for a given EFPD. The burnup levels of different fuel rods at 1420 EFPDs are given in Table 4.

**Table 3**
Distribution of population and power per fuel cycle batch in a typical Westinghouse PWR [33].

| Batch | Number of assemblies | Core fraction vol% ($V_b$) | Relative assembly power ($P_b$) | EFPDs achieved at EOC ($e_b$) |
|---|---|---|---|---|
| 1 | 73 | 38% | 1.25 | 627 |
| 2 | 68 | 35% | 1.19 | 1221 |
| 3 | 52 | 27% | 0.40 | 1420 |
| Total | 193 | 100% | – | – |

On the other hand, St-Aubin and Marleau proposed the $k_{inf}$ averaged over time to evaluate the cycle length without considering neutronic poison [34]. Zhang showed that the critical fuel enrichments determined by the LRM and St-Aubin's method are close [18]. In addition, since the LRM takes the contribution of thermal power at different burnup into account, the present work uses the LRM to determine the critical fuel enrichment. It is noteworthy that $P_b$ of UO$_2$-BeO fuel may be different to the reference values shown in Table 3. Nevertheless, it is possible to optimize the fraction of each batch assembly so that the fraction of total power for each batch $P_b V_b / \sum_b P_b V_b$ is close to the current fuel cladding system. Moreover, the potential influence of different $P_b$ (or $P_b V_b$) values is the critical fuel enrichment, whereas other assembly-level results and conclusion for a given cladding, BeO fraction, and fuel enrichment do not change.

In order to simplify the notation, the volume fraction of BeO in the UO$_2$-BeO fuel is referred to $x$ (in %) and the $^{235}$U enrichment is noted as $y$ (in %) hereinafter. Taking the current UO$_2$-zircaloy fuel-cladding system as the reference, the differences on the reactivity of BeO incorporated fuel and corresponding claddings are calculated by using the LRM. In the present study, $\Delta k_{core}$ is a function of $x$ and $y$. From the analysis of $\Delta k_{core}$ as a function of UO$_2$ content in U$_3$Si$_2$ fuel in Ref. [29], one can suppose that $\Delta k_{core}$ should be approximately a second order polynomial function of the fuel enrichment. Because the quantity of $^{235}$U is proportional to both the fuel enrichment and the fuel volume, we assume that $\Delta k_{core}$ is also a second order polynomial function of the volume of UO$_2$. The term $xy$ is introduced due to the interaction effect between the two variables. $\Delta k_{core}$ is finally expressed by:

$$\Delta k_{core}(x, y) = z + ax + by + cx^2 + dy^2 + fxy, \tag{2}$$

where the coefficients are determined through the least square fitting with the simulated results for each cladding. The coefficient $d$ is expected to be negative because $\Delta k_{core}$ should not increase more quickly with $y$ than linear increment.

The correlation between $x$ and $y$ can be determined via Eq. (2) with $\Delta k_{core}(x, y) = 0$. For a given BeO content $x$, the critical fuel enrichment can be calculated by determining the zeros of Eq. (2):

$$dy^2 + (fx + b)y + (z + ax + cx^2) = 0 \tag{3}$$

It is evident that the UO$_2$-(BeO) fuel without $^{235}$U cannot reach the same cycle length as the reference case. Hence,

$$\Delta k_{core}(x, 0) < 0. \tag{4}$$

The condition Eq. (4) and the existence of physical solution of Eq. (3) implies that Eq. (3) has two positive solutions. The critical fuel enrichment $y_c$ is thus:

$$y_c(x) = [(fx + b) - \sqrt{(fx + b)^2 - 4d(z + ax + cx^2)}]/(-2d). \tag{5}$$

## 3. Results and discussion

### 3.1. Critical fuel enrichment

The infinite multiplication factor $k_{inf}$ is calculated through the Monte Carlo simulation for each case. $\Delta k_{core}$ represents the difference of multiplication factor in the core at the EOC and the values are given in Table 5 for different cases. The visual results of the least square fitting of $\Delta k_{core}$ by using Eq. (2) is shown in Fig. 3 for the zircaloy cladding. The coefficients of Eq. (2) and the corresponding coefficient of determination of least square fitting ($R^2$) are given in Table 6 for the Zr-4, FeCrAl, and SiC claddings. The close to unit values of the coefficient of determination show that the Eq. (2) proposed in the present work can well describe $\Delta k_{core}$ as a function of BeO volume fraction $x$ and the uranium enrichment $y$. In addition, the negative values of $d$ are in accordance with the analysis in Section 2.2.

The critical uranium enrichments for different claddings and different BeO contents can be determined by using Eq. (5). An important remark is the sensitivity of $\Delta k_{core}$ to the fuel enrichment $y$:

$$\partial \Delta k_{core}(x, y)/\partial y = 2dy + (fx + b). \tag{6}$$

For a BeO volume fraction $x > 5\%$ and an uranium enrichment $y < 8\%$ (which is the general case of the critical uranium enrichments $y_c$ for $x < 36.4\%$), $\partial \Delta k_{core}(x, 0)/\partial y > 0.03$ , which means 0.1% uranium enrichment has more than 300 pcm influence on $\Delta k_{core}$ . Neutronic verification of the critical fuel enrichment $y_c$ obtained by Eq. (5) is suggested because of the large sensitivity of $\Delta k_{core}$ to fuel enrichment.

In Table 7, the critical uranium enrichments calculated through Eq. (5) are validated against the Monte Carlo simulations, where the two results are very close to each other for most cases. The only two exceptions are the FeCrAl cladded UO$_2$-(5vol%)BeO and UO$_2$-(10vol%)BeO fuels, of which the solutions of Eq. (5) are listed in parenthesis and have 0.07% absolute differences with the Monte Carlo simulations. Generally speaking, a larger cross section of claddings needs larger critical uranium enrichment. The present results shown in Table 5 are in agreement with such statement. The thermal neutron absorption cross sections of claddings are 0.20 barns, 2.43 barns, and 0.086 barns for the Zr-4, FeCrAl, and SiC claddings, respectively [35]. Among all content fractions of BeO, the fuels with FeCrAl and SiC claddings have the largest and smallest critical uranium enrichment, respectively.

The $k_{inf}$ for the reference case and for the UO$_2$-(xvol%)BeO fuel with $x = 10$ and 36.4 combined with the FeCrAl, zircaloy, and SiC claddings are shown in Fig. 4. Similar $k_{inf}$ are observed for the zircaloy alloy cladding and SiC cladding. Slightly higher values at low burnup for the SiC cladding is due to the lower thermal neutron absorption cross sections. The results also point out the large influence of large thermal neutron absorption cross section of the FeCrAl on the $k_{inf}$.

For fuel incorporated with other elements (such as BeO in the present work), a widely used method to roughly determine the critical uranium enrichment is the guaranty of the $^{235}$U loading in the fuel. The present work compares critical fuel enrichments determined by this method (referred to $y_U$ hereinafter) and by the LRM (i.e. $y_c$). The values and the differences between $y_U$ and $y_c$ obtained from Eq. (5) are presented in Table 8 for the Zr-4 cladding. The corresponding $\Delta k_{core}(x, y_U)$ can be calculated directly through Eq. (6) with $\Delta y = y_U - y_c$ for each BeO volume fraction $x$ because $\Delta k_{core}(x, y_c) = 0$. Since the calculation of $y_U$ requires a known $^{235}$U enrichment for each cladding case, the present work assumes $y_U = y_c$ for the UO$_2$-(5vol%)BeO fuel in the analyses of $y_U$ for the FeCrAl and SiC claddings. The differences $\Delta y = y_U - y_c$ with the FeCrAl and SiC claddings are given in Table 9.

Results in Table 9 show that the method to keep the $^{235}$U loading for the SiC cladding corresponds well to the results obtained by Eq. (5), which is based on the LRM. The difference $\Delta y = y_U - y_c$ increases with $x$ for the zircaloy and FeCrAl claddings. The discrepancies of the uranium enrichment calculated with the two methods are within 0.12% for BeO volume fraction less than 36.4% with the zircaloy cladding. However, large discrepancies are observed for the FeCrAl cladding. The discrepancies in critical enrichments determined by the two different methods for the different claddings can be caused by two reasons:

- The first one is the neutron absorption cross section of the cladding, which has a similar role as BeO of the negative contribution to reactivity. An increase of BeO in the fuel has thus less effect on the reactivity for the cladding with higher thermal absorption cross

**Table 4**
Equivalent burnup (in MWd/kgU) of different UO$_2$-BeO fuels at 1420 EFPDs.

| Cladding | Volume fraction of BeO in UO$_2$-BeO fuel | | | | | |
| --- | --- | --- | --- | --- | --- | --- |
| | 0% | 5% | 10% | 20% | 30% | 36.4% |
| Zr-4 and SiC | 54.4 | 57.3 | 60.5 | 68.0 | 77.8 | 85.6 |
| FeCrAl | 49.2 | 51.8 | 54.6 | 61.5 | 70.2 | 77.3 |

**Table 5**

$\Delta k_{core}$ between the investigated cases and the reference case.

| Cladding material | Enrichment $y$ (%) | Volume fraction of BeO in UO$_2$-BeO fuel $x$ (%) | | | | |
|---|---|---|---|---|---|---|
| | | 5 | 10 | 20 | 30 | 36.4 |
| Zr-4 | 4.9 | −0.0141 | −0.0305 | −0.0713 | −0.1243 | −0.1645 |
| | 6.0 | 0.0472 | 0.0341 | −0.0026 | −0.0537 | −0.0975 |
| | 7.0 | 0.0949 | 0.0828 | 0.0519 | 0.0057 | −0.0359 |
| | 8.0 | – | 0.1230 | 0.0968 | 0.0583 | 0.0207 |
| | 10.0 | – | 0.1859 | 0.1671 | 0.1384 | 0.1120 |
| FeCrAl | 4.9 | −0.0372 | −0.0387 | −0.0746 | −0.1235 | −0.1637 |
| | 6.0 | 0.0209 | 0.0216 | −0.0105 | −0.0559 | −0.0949 |
| | 7.0 | 0.0657 | 0.0674 | 0.0390 | −0.0008 | −0.0363 |
| | 8.0 | – | 0.1065 | 0.0813 | 0.0471 | 0.0150 |
| | 10.0 | – | 0.1680 | 0.1490 | 0.1216 | 0.0982 |
| SiC | 4.9 | −0.0067 | −0.0235 | −0.0639 | −0.1176 | −0.1588 |
| | 6.0 | 0.0547 | 0.0410 | 0.0048 | −0.0467 | −0.0907 |
| | 7.0 | 0.1014 | 0.0896 | 0.0592 | 0.0136 | −0.0292 |
| | 8.0 | – | 0.1302 | 0.1045 | 0.0652 | 0.0289 |
| | 10.0 | – | 0.1932 | 0.1932 | 0.1461 | 0.1198 |

section, and subsequently less additional enrichment is required for higher BeO contents. While the LRM model, which is based on neutron physical calculations, accounts for this effect, the simplistic approach of constant $^{235}$U loading is not able to take such effects into consideration.

- Another reason is the increase of the moderator to fuel (uranium) ratio with the content of BeO. The reactivity increases but the increment rate decreases with the moderator to fuel ratio in the range that the present work based on. Since the 350 μm FeCrAl cladding leads to about 11% higher fuel volume compared to the cases with Zr-4 and SiC claddings, the moderator to fuel ratio for the case with FeCrAl is smaller than the cases with Zr-4 and SiC. As a consequence, an increase of the moderator to fuel ratio has a more evident effect for the case with FeCrAl cladding compared to the cases with Zr-4 and SiC claddings.

### 3.2. Gas production

As explained previously, the (n,α) cross sections of $^9$Be in the neutron energy range from 700 keV up to 3 MeV are relatively large (as shown in Fig. 2). The production of $^4$He should be considerable in the UO$_2$-BeO fuel. It should be noted again that only the neutronic analyses are performed in our simulations, while the diffusion of gas is not

**Table 6**

Coefficients and corresponding uncertainties of Eq. (2) obtained by least square fitting. $R^2$ is the coefficient of determination of least square fitting.

| Cladding | Zr-4 | | FeCrAl | | SiC | |
|---|---|---|---|---|---|---|
| $z$ | −0.35467 | 2.7% | −0.37151 | 3.3% | −0.33910 | 4.6% |
| $a$ | −0.00433 | 6.0% | −0.00241 | 13% | −0.00386 | 10% |
| $b$ | 0.08918 | 2.9% | 0.08620 | 3.8% | 0.08517 | 4.9% |
| $c$ | −7.11E-05 | 7.5% | −8.70E-05 | 7.7% | −8.18E-05 | 10% |
| $d$ | −0.00347 | 4.9% | −0.00330 | 6.5% | −0.00313 | 8.8% |
| $f$ | 4.718E-04 | 5.7% | 3.795E-04 | 8.9% | 4.652E-04 | 9.3% |
| $R^2$ | 0.9994 | | 0.9989 | | 0.9985 | |

**Table 7**

Critical uranium enrichment $y_c$ (in %) to keep the cycle length for different BeO content with Zr-4, FeCrAl, and SiC claddings.

| $x$ (%) | 5 | 10 | 20 | 30 | 36.4 |
|---|---|---|---|---|---|
| Zr-4 | 5.13 | 5.39 | 6.05 | 6.91 | 7.58 |
| FeCrAl | 5.59 (5.52)* | 5.60 (5.67)* | 6.20 | 7.01 | 7.70 |
| SiC | 5.02 | 5.26 | 5.91 | 6.79 | 7.47 |

* The two values in parenthesis are solutions of Eq. (5) but Monte Carlo simulations show better $\Delta k_{core}$ with the other two values.

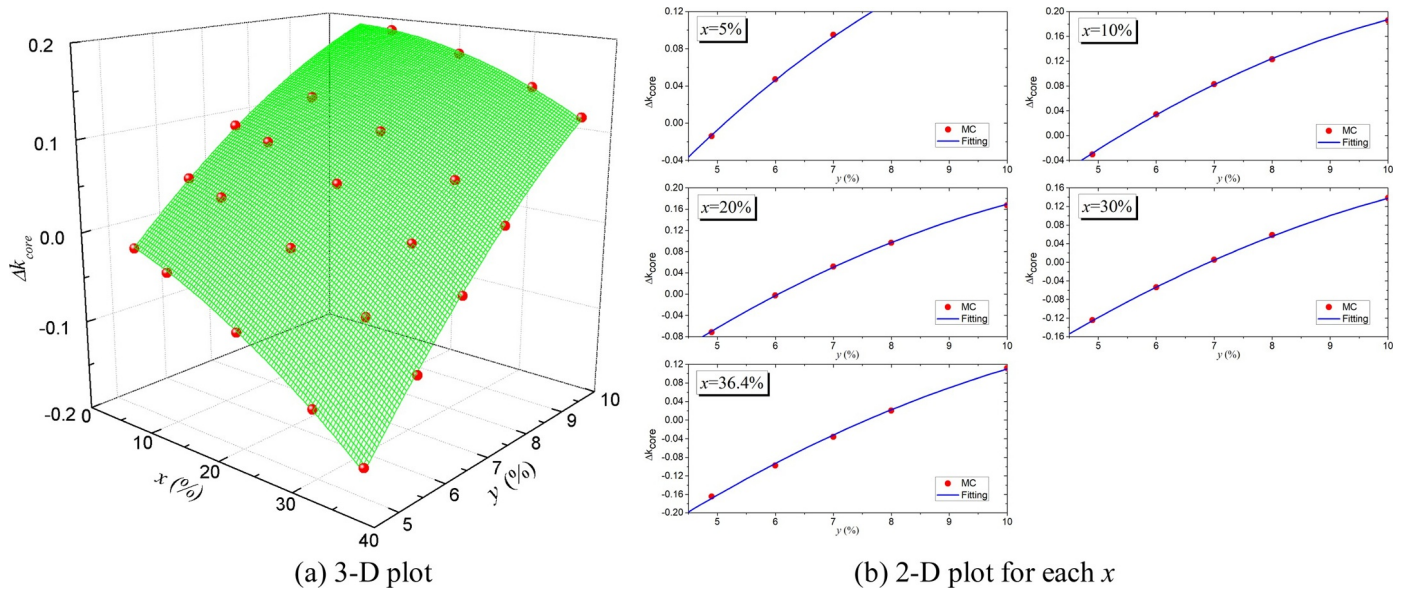

(a) 3-D plot  (b) 2-D plot for each $x$

**Fig. 3.** Fitting results of $\Delta k_{core}$ for the Zr-4 cladding. Red points are Monte Carlo simulation results and the green surface is the least square fitting of Eq. (2).

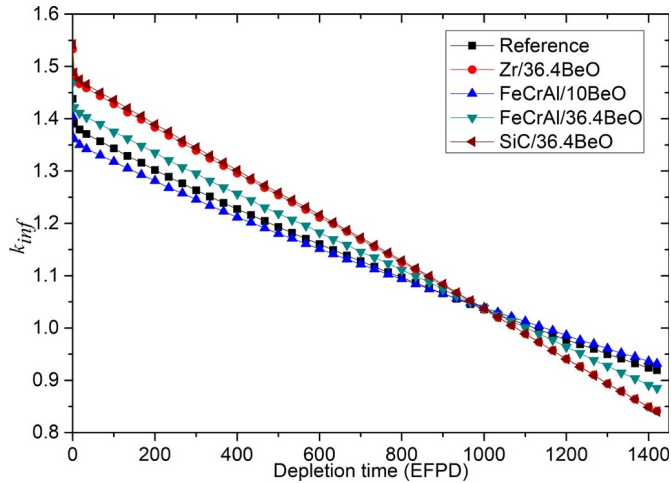

**Fig. 4.** Infinite multiplication factor $k_{inf}$ versus effective depletion time for the reference case and the $UO_2$-($x$vol%)BeO ($x$ = 10, 36.4) fuel combined with the different claddings.

**Table 8**

Critical uranium enrichment in $UO_2$-($x$vol%)BeO fuel by keeping the $^{235}U$ load ($y_U$) and comparison with $y_c$ with zircaloy alloy cladding.

| $x$ (%) | 5 | 10 | 20 | 30 | 36.4 |
|---|---|---|---|---|---|
| $y_U$ (%) | 5.16 | 5.44 | 6.13 | 7.00 | 7.70 |
| $y_U$- $y_c$ (%) | 0.03 | 0.05 | 0.08 | 0.09 | 0.12 |
| $\Delta k_{core}(x, y_U)$ | 0.0016 | 0.0031 | 0.0042 | 0.0050 | 0.0066 |

**Table 9**

Difference of critical uranium enrichment in $UO_2$-($x$vol%)BeO fuel obtained by keeping the $^{235}U$ load ($y_U$) and by using Eq. (5) ($y_c$) (i.e. $y_U$ - $y_c$). $y_U$ = $y_c$ is assumed for $UO_2$-(5vol%)BeO.

| $x$ (%) | 5 | 10 | 20 | 30 | 36.4 |
|---|---|---|---|---|---|
| FeCrAl | – | 0.15 | 0.36 | 0.48 | 0.55 |
| SiC | – | 0.04 | 0.05 | 0.03 | 0.03 |

considered. Fig. 5 represents the atomic density of $^4$He in the fuel for the reference case and six combinations of $UO_2$-BeO fuel with different claddings. As expected, the quantity of $^4$He in the $UO_2$-BeO fuel is quite larger than that produced in the current $UO_2$ fuel.

The first row in Table 10 lists the atomic density of $^4$He for several cases shown in Fig. 5 at the End of Life (EOL). The $^4$He production in the $UO_2$-(36.4vol%)BeO fuel is 15 times larger than the current $UO_2$ fuel. As a result, although the additional BeO in $UO_2$ fuel improves the thermal conductivity, the higher $^4$He production should be paid attention to.

Except $^4$He, other gasses are produced through the fission reaction. The main fission gasses are isotopes of xenon (Xe) and krypton (Kr). Fig. 6 shows the sum of atomic density of all isotopes of Kr and Xe as a function of fuel exposure for different cases. Results point out lower Kr and Xe productions in the case of FeCrAl cladding than those in the zircaloy alloy and SiC claddings. The main reason is the larger fuel volume owing to the thinner cladding thickness. The difference between zircaloy cladding and SiC cladding is not evident because of the same fuel quantity and the quite small thermal neutron absorption cross sections of zircaloy and SiC. Results show that the effect of BeO loading on the production of Kr and Xe is not important. Table 10 summarizes the production of $^4$He, Kr, and Xe at the EOL.

Due to the similar production of Kr and Xe in the cases of zircaloy and SiC claddings to the reference case, $^4$He produced through the (n,$\alpha$) reaction of $^9$Be has a direct effect on the total gas production. The total gas density in the fuel with FeCrAl cladding is lower than that with the Zr-4 and SiC cladding due to the larger fuel volume. A general conclusion is that $^4$He has almost 1% contribution to the total gas production in the reference case, while 20% contribution to the total gas production is found in the $UO_2$-(36.4vol%)BeO fuel and 7% is found in $UO_2$-(10vol%)BeO. The last row in Table 10, the ratio of total gas production to that in the reference case, illustrates the increase of gas release for the different cases. The increments of the total gas release are within 23%.

It is noteworthy that Fig. 6 and Table 10 show more gas production in $UO_2$-BeO fuel than the reference case at a specific EFPD, while Ref. [17] find less fission gas release in $UO_2$-BeO fuel at a given burnup level. One reason is the lower fuel temperature by incorporating BeO in the $UO_2$ fuel. In addition, the change of microstructure leads to the difference between gas production and gas release. A remarkable reason is the use of different units for fuel exposure. For a given EFPD, more BeO incorporation implies less $UO_2$ that signifies higher burnup. For example, comparing the 36.4%BeO (i.e. 13.6wt%BeO) incorporation with the reference case, the same EFPD leads to a higher burnup of the $UO_2$-36.4%BeO fuel by a factor of 1.16. The relationship between EFPD and burnup is given in Section 2 and summarized in Table 4.

### 3.3. Relative power distribution in an assembly

The distribution of power in a fuel assembly is displayed in Fig. 7 for the reference $UO_2$-zircaloy case at the BOL. Higher power is observed for the fuel pins in the vicinity of the guide tubes due to the higher local moderator-to-fuel ratio. The Power Peaking Factor (PPF) is defined as the ratio of the maximum power to average power in the fuel assembly. The PPFs in the cases of zircaloy cladding with different BeO volume fractions are illustrated in Fig. 8. Results in Fig. 8 indicate the decrease of the PPF with BeO content at high burnup. In other words, the $UO_2$ fuel with a higher BeO percentage has flatter power distribution in an assembly.

Fig. 9 represents the PPF for the $UO_2$-(36.4vol%)BeO fuel with different claddings. The PPFs are almost the same for the zircaloy and SiC claddings. The PPF values for the case of FeCrAl cladding are larger than the other cases. Therefore, the use of FeCrAl cladding may lead to smaller margins linked to the hotspots, such as the melting of the fuel and the evaporation of the coolant.

### 3.4. Radial power distributions

The investigations on the radial distribution of power are important

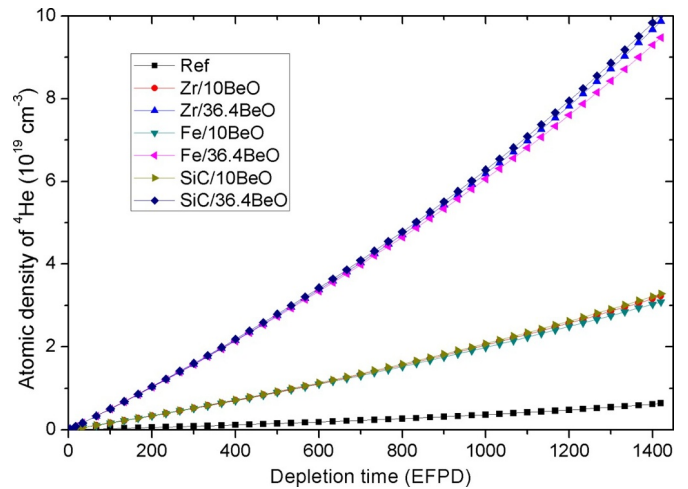

**Fig. 5.** Atomic density of $^4$He produced in the fuel for different cases with the same cycle length, including the reference case and the $UO_2$-($x$vol%)BeO ($x$ = 10, 36.4) fuel combined with the zircaloy, FeCrAl, and SiC claddings.

**Table 10**

Atomic density (in $10^{19}$ cm$^{-3}$) of $^4$He, Kr, and Xe for the reference case and the $UO_2$-($x$vol%)BeO ($x = 10$ or 36.4) fuel combined with different claddings at the EOL. The last row gives the ratio of total gas production to that in the reference case.

| Case[a] | Ref. | Zr/10 | Zr/36.4 | FeCrAl/36.4 | SiC/36.4 |
|---|---|---|---|---|---|
| $^4$He | 0.64 | 3.22 | 9.87 | 9.48 (10.5)[b] | 10.0 |
| Kr | 3.92 | 3.91 | 3.99 | 3.66 (4.07) | 3.97 |
| Xe | 36.7 | 36.4 | 36.7 | 32.6 (36.2) | 36.7 |
| Total[c] | 41.2 | 43.6 | 50.6 | 45.7 (50.8) | 50.7 |
| Ratio | 1.00 | 1.06 | 1.23 | 1.11 (1.23) | 1.23 |

[a] Zr and Fe respectively represent the Zr-4 and FeCrAl claddings; 10 and 36.4 stand for the volume fraction of BeO in the fuel.

[b] Concentrations in parenthesis are produced gas divided by the fuel volume with 572 μm thick cladding (i.e. 1.107 shown in Table 2).

[c] "Total" shows the total atomic density of the three atomic densities.

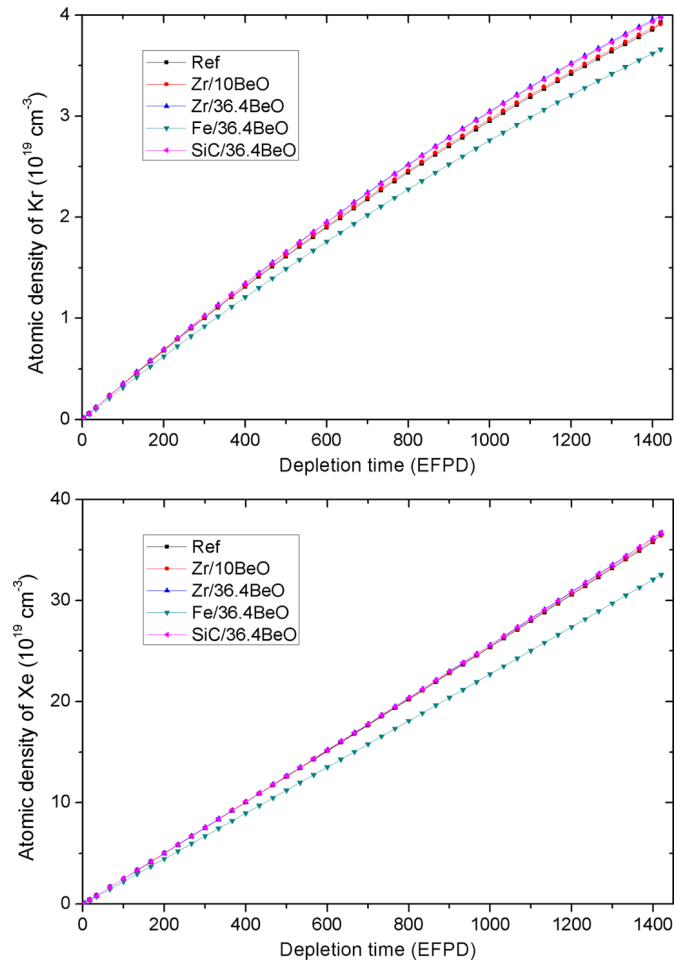

**Fig. 6.** Sum of atomic density of all isotopes of Kr (upper) and Xe (lower) produced in the fuel for different case with the same cycle length, including the reference case, and the $UO_2$-($x$vol%)BeO ($x = 10$, 36.4) fuel combined with the zircaloy, and the $UO_2$-(36.4vol%)BeO fuel combined with the FeCrAl and SiC claddings.

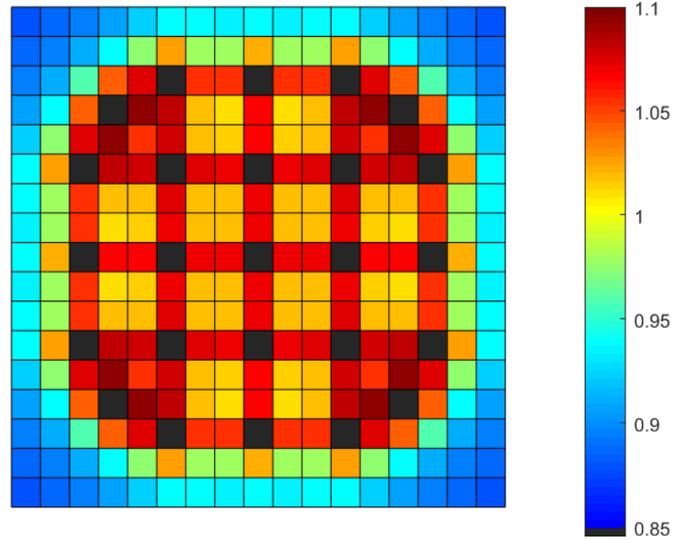

**Fig. 7.** Relative power distribution at the BOL for the reference $UO_2$-Zircaloy case. The black points out the guide tubes.

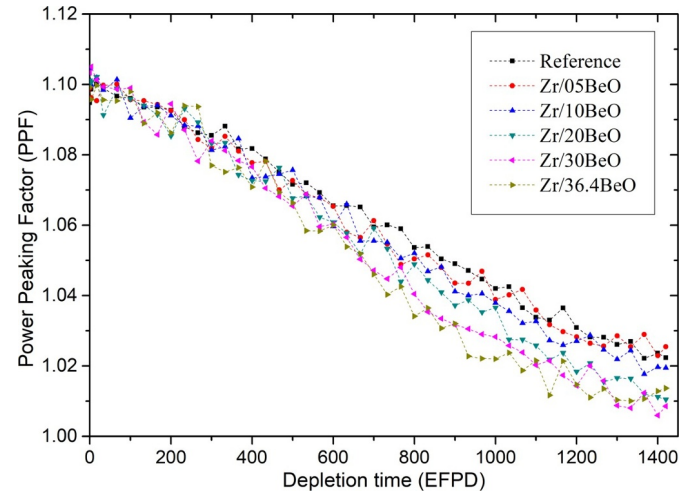

**Fig. 8.** Power peaking factor for different BeO volume fraction with Zr-4 cladding.

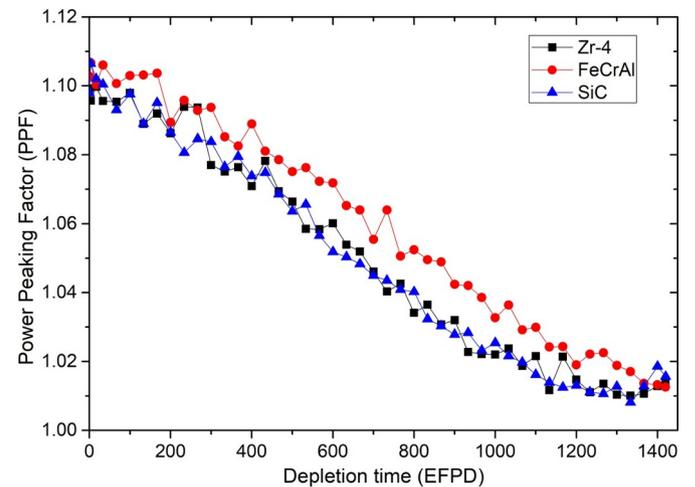

**Fig. 9.** Power peaking factor for $UO_2$-(36.4vol%)BeO fuel with different claddings.

because the power distribution can give feedback to the neutronic calculations and the multi-physics coupling study. Fig. 10 shows the radial relative power distributions for the Zr-4 cladding with $UO_2$-($x$vol%)BeO ($x = 10$, 36.4) fuel at the BOL, MOL, and EOL. The radial power distributions are quite similar at the BOL. However, Monte Carlo simulations show flatter radial power distributions in the higher BeO content $UO_2$-BeO fuel after the BOL. Lower relative power at the periphery is due to the lower concentration of $^{235}$U and $^{239}$Pu (Figs. 14 and

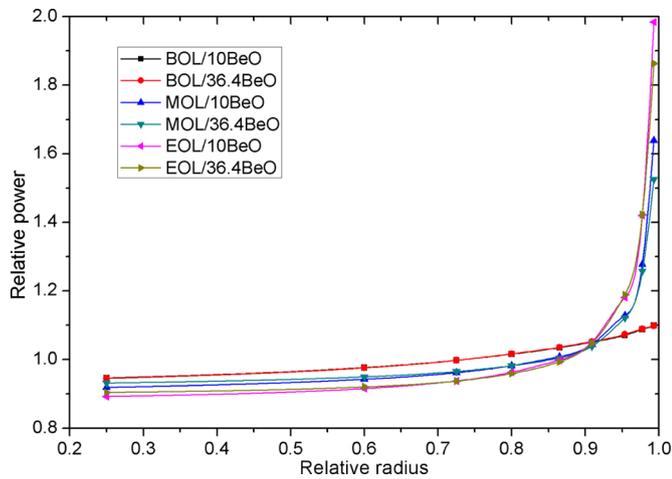

**Fig. 10.** Radial power distribution in the UO$_2$-(xvol%)BeO ($x$ = 10, 36.4) fuel combined with the zircaloy cladding at the BOL, MOL, and EOL.

15) near the surface of the fuel pellet. The flatter radial power distribution is an advantage of the BeO incorporated fuel from the consideration of neutronic properties, such as the more uniform radial burnup distribution. The radial profile of thermal power can be further flattened if a dual-coolant annular design is used for fuel rod [36].

Fig. 11. presents the radial distribution of normalized power in the UO$_2$-(36.4vol%)BeO fuel combined with the zircaloy, FeCrAl, and SiC claddings at the MOL and EOL. Comparing with the results in Fig. 10, the influence of the cladding on the radial power distribution is negligible. This conclusion is in agreement with the results obtained in Ref. [29], which shows the similar radial power distribution for the FeCrAl cladded U$_3$Si$_2$ and U$_3$Si$_2$/UO$_2$ fuels and the current zircaloy cladded UO$_2$ fuel. The similarity between radial power distribution with the FeCrAl cladding and those with other claddings is the compensated effect of higher relative concentration of fissile isotopes at periphery (such as $^{235}$U and $^{239}$Pu in Figs. 14 and 15, which has positive contribution to the sharp power distribution at periphery) and the higher thermal neutron absorption cross sections (which weaken the effect induced by fissile isotopes). An intuitive explication is the more accumulated neutron absorption isotopes due to the lower percentage of the thermal neutron. Further investigation should be performed to explain the similar radial power distribution while the concentrations of fissile isotopes are different.

### 3.5. Isotopic concentrations

In an uranium fueled PWR, the two most important isotopes for fission are $^{235}$U and $^{239}$Pu. The concentrations of $^{235}$U (and $^{239}$Pu) for the reference case, the UO$_2$-(xvol%)BeO (with $x$ = 10, 36.4) fuel combined with the zircaloy cladding, and the UO$_2$-(36.4vol%)BeO fuel with FeCrAl and SiC claddings are plotted in Fig. 12 (and Fig. 13) as a function of the fuel exposure. In Fig. 12, the initial concentrations (given in the caption) are close for different scenarios because a higher BeO fraction requires a higher $^{235}$U enrichment (see Table 7) to keep the same cycle length.

The concentration of $^{235}$U decreases with the volume fraction of BeO. This phenomenon is mainly induced by the much lower $^{238}$U quantity in the BeO incorporated fuel, which leads to a lower concentration of $^{239}$Pu as shown in Fig. 13. Due to the less contribution of $^{239}$Pu to the power, more fission of $^{235}$U is required. Accordingly, the concentration of $^{235}$U decreases more quickly with effective full depletion time for higher BeO content in the fuel. A secondary-order reason is the lower initial concentration of $^{235}$U in higher BeO content fuel, which is shown in Tables 8 and Table 9, and Fig. 12. For the same BeO volume fraction, the case of FeCrAl cladding has a higher $^{239}$Pu concentration than that in the zircaloy cladding case because of the hardening of neutron spectrum [4]. Slightly lower $^{239}$Pu concentration with the SiC cladding is due to lower thermal neutron absorption cross section of SiC than that of Zr-4, which leads to the softening of neutron spectrum.

The investigations on the radial distribution of isotopic concentrations of major actinides help us to understand the radial power distribution. In addition, it can provide information to the neutronic calculations. The principal fissile isotopes in an uranium fueled PWR are $^{235}$U, $^{239}$Pu, and $^{241}$Pu. The relative concentrations of $^{235}$U (and $^{239}$Pu) at the EOL are shown in Fig. 14 (and Fig. 15) for the UO$_2$-(xvol%)BeO ($x$ = 10, 36.4) fuel combined with the zircaloy cladding, and the UO$_2$-(36.4vol%)BeO fuel with FeCrAl and SiC claddings.

Fig. 14. shows that the radial profiles of $^{235}$U concentrations are relatively flat until EOL. The radial distributions of $^{235}$U concentrations with BeO incorporation are not so different to those of the reference case and three U$_3$Si$_2$-FeCrAl combinations with the same cycle length shown in Ref. [29]. The discrepancies among different cases are the results of different average concentrations, different neutron flux spectra, different compositions of fuels, etc. Nevertheless, the relative difference on the radial profiles of $^{235}$U concentrations quite small (< 6% for all cases until EOL) for all cases studied in the present work.

$^{239}$Pu is produced by a capture reaction of $^{238}$U and two times of $\beta$ decay after. In PWRs, the increase of thermal neutron density with radius in the fuel pellet is the main reason for the sharp radial profile of $^{239}$Pu concentration. Comparing the cases of zircaloy cladding, one can find a sharper radial distribution of $^{239}$Pu for a lower BeO content fuel from Fig. 15. Fig. 15 also shows that the radial distribution of $^{239}$Pu depends on the material of cladding. Similar to the conclusion on $^{235}$U concentration, the radial distributions of $^{239}$Pu concentrations with BeO incorporation are close to those of the reference case and three U$_3$Si$_2$-FeCrAl combinations with the same cycle length shown in Ref. [29]. The discrepancies on the radial profiles of $^{239}$Pu concentrations among cases of different claddings are within 8% for different combinations considered in the present work.

### 4. Conclusions

Neutron physical calculations are performed on different ATF candidate fuels based on UO$_2$ with incorporation of different volume fractions of BeO and different claddings for a typical PWR using the Monte Carlo method. Application of the Linear Reactivity Model shows that the multiplication factor at the EOC ($\Delta k_{core}$) can be well described by a second order polynomial function of the BeO volume fraction ($x$) and the uranium enrichment ($y$). The critical fuel enrichment is determined for each volume fraction of BeO with the Zr-4, FeCrAl, and SiC claddings. Similar infinite multiplication factors $k_{inf}$ are observed for the zircaloy alloy cladding and SiC cladding. Slightly higher $k_{inf}$ with the SiC cladding at low burnup is due to the lower thermal neutron absorption cross sections of SiC. The results point out the lower $k_{inf}$ with the FeCrAl cladding induced by the larger thermal neutron absorption cross section of FeCrAl.

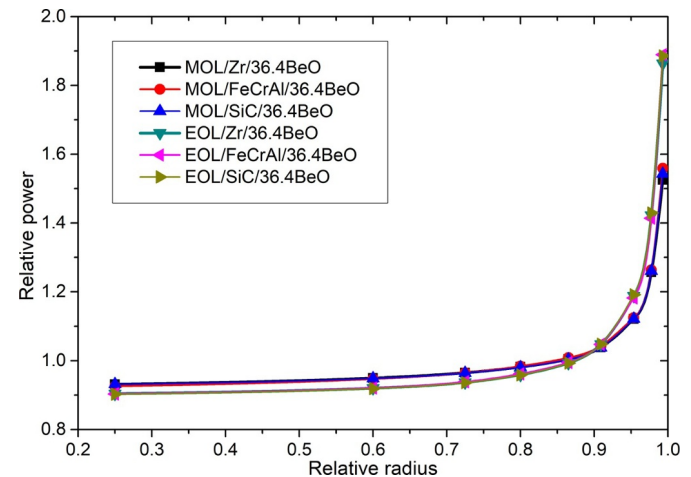

**Fig. 11.** Radial power distribution in the UO$_2$-(36.4vol%)BeO fuel combined with the zircaloy, FeCrAl, and SiC claddings at the MOL and EOL.

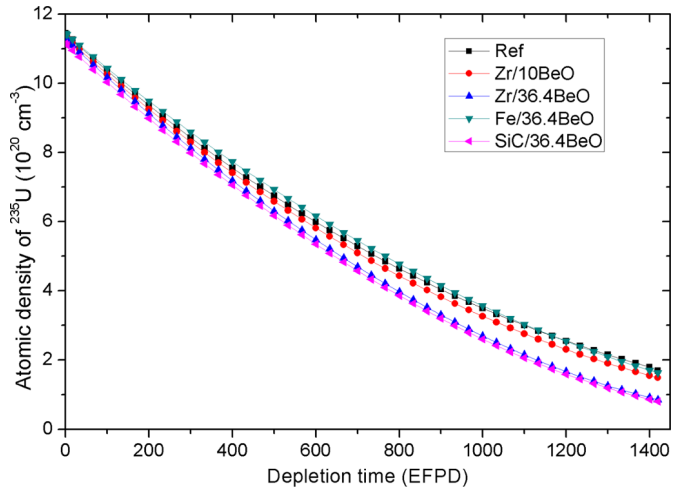

**Fig. 12.** Atomic density of $^{235}$U in the fuel for different cases with the same cycle length, including the reference case, the UO$_2$-($x$vol%)BeO ($x$ = 10, 36.4) fuel combined with the zircaloy cladding, and the UO$_2$-(36.4vol%)BeO fuel with FeCrAl and SiC claddings. The initial concentrations are 11.45, 11.36, 11.29, 11.44, and 11.15 × 10$^{20}$ cm$^{-3}$ for the cases orderly indicated by the legend.

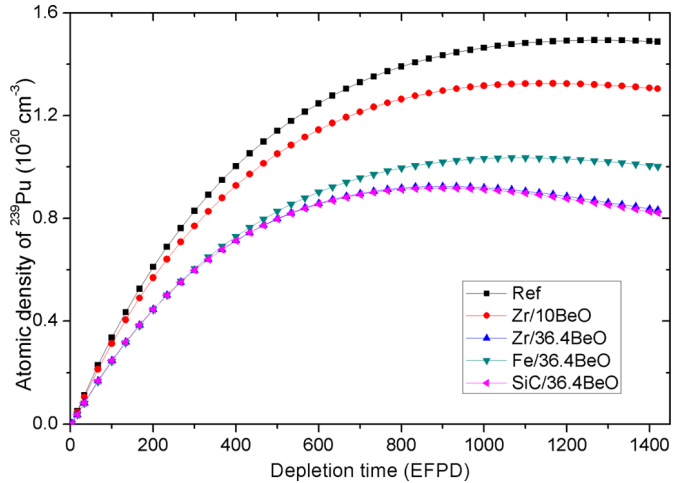

**Fig. 13.** Atomic density of $^{239}$Pu in the fuel for different cases with the same cycle length, including the reference case, the UO$_2$-($x$vol%)BeO ($x$ = 10, 36.4) fuel combined with the zircaloy cladding, and the UO$_2$-(36.4vol%)BeO fuel with FeCrAl and SiC claddings.

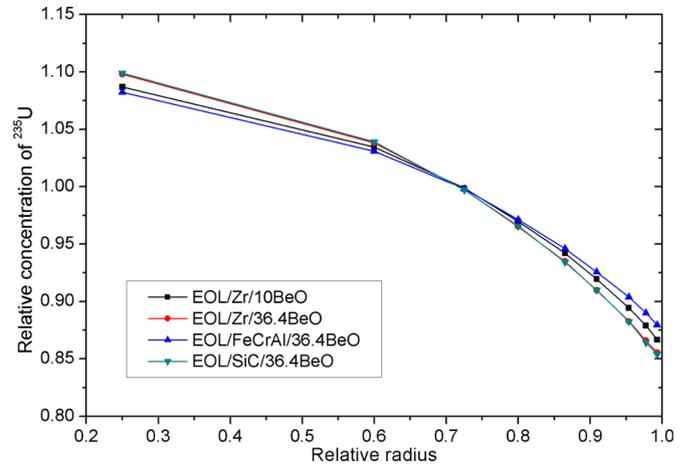

**Fig. 14.** Relative concentrations of $^{235}$U in the fuel for different cases with the same cycle length, including the UO$_2$-($x$vol%)BeO ($x$ = 10, 36.4) fuel combined with the zircaloy cladding, and the UO$_2$-(36.4vol%)BeO fuel with FeCrAl and SiC claddings.

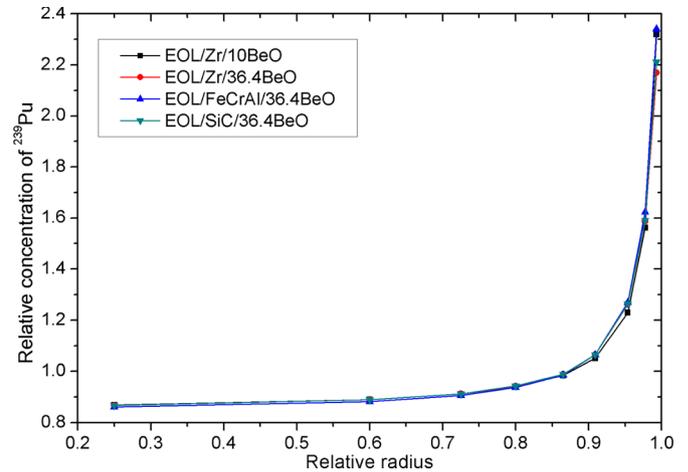

**Fig. 15.** Relative concentrations of $^{239}$Pu in the fuel for different cases with the same cycle length, including the UO$_2$-($x$vol%)BeO ($x$ = 10, 36.4) fuel combined with the zircaloy cladding, and the UO$_2$-(36.4vol%)BeO fuel with FeCrAl and SiC claddings.

Compared with the current UO$_2$ fuel, much more $^4$He is produced in the UO$_2$-BeO fuel due to the relatively large (n,$\alpha$) cross section of the $^9$Be in the neutron energy range from 700 keV up to 3 MeV. The production of krypton and xenon in the UO$_2$-BeO-zircaloy and UO$_2$-BeO-SiC fuel–cladding systems are similar to those in the current UO$_2$-zircaloy system, while those in the UO$_2$-BeO-FeCrAl system are less than the current fuel-cladding combination. Taking the current UO$_2$-zircaloy fuel-cladding combination as the reference, 23% more total gas production is obtained in the UO$_2$-(36.4vol%)BeO fuel with the Zr-4 and SiC claddings (FeCrAl cladding) at the EOL, while 11% more total gas production is found with the FeCrAl cladding.

The peak power in an assembly decreases with the BeO content in the UO$_2$-BeO fuel at the high burnup level. For the same BeO content, such as UO$_2$-(36.4vol%)BeO, the peak power is quite similar for the cases with the zircaloy and SiC claddings, while that with the FeCrAl cladding is higher. The radial power distribution in the UO$_2$-BeO fuel is not sensitive to the claddings considered in the present work. Quite similar radial power distribution is observed at the BOL for the different BeO fractions. After the BOL, a flatter radial power distribution is shown for a higher BeO fraction in the UO$_2$-BeO fuel.

The concentration of $^{239}$Pu is lower in a higher BeO content fuel because of less $^{238}$U in the fuel. $^{235}$U concentration is also lower for a higher BeO fraction due to less $^{239}$Pu and slightly low initial concentration. The concentrations of $^{235}$U and $^{239}$Pu are similar between the zircaloy cladding and SiC cladding cases. The higher concentration of $^{239}$Pu with the FeCrAl cladding is mainly due to the higher initial concentration of $^{238}$U and harder neutron spectrum. For the radial profiles of isotopic concentrations, simulation results show that the radial distributions of the relative concentrations of $^{235}$U and $^{239}$Pu are quite similar among different BeO fractions and different claddings for each EFPD.

## CRediT authorship contribution statement

**Shengli Chen:** Conceptualization, Data curation, Formal analysis, Investigation, Methodology, Software, Validation, Visualization, Writing - original draft. **Cenxi Yuan:** Formal analysis, Investigation, Methodology, Supervision, Validation, Visualization, Writing - original draft, Funding acquisition.

## Declaration of Competing Interest

The authors declare that they have no known compating financial interest or personal relationships that could have appeared to influence the work reported in this paper.

## Acknowledgments

The authors acknowledge the authorized usage of the RMC code from Tsinghua University for this study. This work has been supported by the National Key Research and Development Program of China under Grant No. 2018YFB1900405, the National Natural Science Foundation of China under Grant No. 11775316, the Tip-top Scientific and Technical Innovative Youth Talents of Guangdong special support program under Grant No. 2016TQ03N575, the Science and Technology Planning Project of Guangdong Province, China under Grant No. 2019A050510022, and the computational resources from Sun Yat-sen University and National Supercomputer Center in Guangzhou.